\documentstyle[12pt]{article}
\begin{document}
\title{A Note on Classical Solution of Chaplygin-gas as d-Branes}
\author{Naohisa Ogawa 
\thanks{Email: ogawa@particle.sci.hokudai.ac.jp}
\\Department of Physics, Faculty of Science, 
\\Hokkaido University, Sapporo 060-0810 Japan}
\date{March,  2000}
\maketitle

\begin{abstract}
The classical solution of bosonic d-brane in (d+1,1) space-time is studied. 
We work with light-cone gauge and reduce the problem into Chaplygin gas problem.
The static equation is equivalent to vanishing of extrinsic mean curvature, 
which is similar to Einstein equation in vacuum.
We show that the d-brane problem in this gauge is closely related to Plateau problem,
and we give some non-trivial solutions from minimal surfaces. 
The solutions of d-1,d,d+1 spatial dimensions are obtained 
from d-dimensional minimal surfaces as solutions of Plateau problem.
In addition we discuss on the relation to Hamiltonian-BRST formalism for d-branes.
\end{abstract}

\section{Introduction}
~~~~ The theory of d-brane in (d+1,1) space-time is equivalent to the fluid dynamics.
This is first found by  Goldstone, and developed by Hoppe and Bordemann \cite{Goldstone}, \cite{Hoppe}.
 Starting from Nambu-Goto action, they fixed the time reparametrization by using light-cone gauge and 
 solved the momentum constraint in miracle way with some gauge condition. 
The action obtained in this way gives d-dimensional irrotational Euler equation with given pressure as
\begin{equation}
\dot{\bf v} + {\bf v} \cdot \nabla {\bf v} = - \frac{\nabla P}{\rho}, ~~
P = -\frac{2 \lambda}{\rho}, ~~ {\bf v} =\nabla \theta,
\end{equation}
where $\rho$ is the matter density,  $\lambda$ is some constant, and $\theta$ is velocity potential. 
The action has the form.
\begin{equation}
S_C=\int d^{d+1}r [\theta \dot{\rho} - (\rho \frac{(\nabla \theta)^2}{2m} + \frac{\lambda}{\rho})]. \label{eqn:chap}
\end{equation}

Due to the last term we can eliminate $\rho$ variable from the theory by using equation of motion 
if we take $\lambda \neq 0$.
Then we have 
\begin{equation}
S_C=-2\sqrt{\lambda} \int d^{d+1}r [\dot{\theta} + \frac{(\nabla \theta)^2}{2m}]^{1/2}.\label{eqn:chaplygin}
\end{equation}
This kind of fluid is called Chaplygin gas (see the reference of \cite{Jackiw-Polychronakos}),
and it has d+2 dimensional Poincar\'e symmetry with each generators given by Hoppe \cite{Hoppe}. 
The similar discussion is also given in 4-dimensional scalar field theory 
using light-front formalism given by Jevicki \cite{Jevicki}.
For the theory of fluid, time and space translation (energy and momentum conservation), 
rotation (angular momentum conservation), Galileo boost and phase symmetry (matter conservation) 
are the natural symmetries. But the number of these generators is totally  $(d^2 +3d +4)/2$ and 
is missing $d+1$ generators to construct $d+2$ dimensional Poincar\'e generators.
These d+1 generators are the hidden symmetries in the theory of fluid dynamics, 
and this point was made clear by Bazeia and Jackiw.
One is the time-rescaling symmetry (dilatation), and another d-generators induce 
the field dependent transformation 
(Hereafter abbreviated as FDT) which mixes the dynamical field with space-time  \cite{Jackiw-Bazeia}.
On the other hand, Hoppe \cite{Hoppe} has also shown another gauge fixing 
(we call Cartesian parametrization hereafter) which reduces the Nambu-Goto theory to Poincar\'e invariant 
Born-Infeld model given by
\begin{equation}
S_{BI} = \int d^{d+1} r [\theta \dot{\rho} - \sqrt{\rho^2 c^2 + a^2} \sqrt{m^2 c^2+ (\nabla \theta)^2}],
\end{equation}
or its another representation
\begin{equation}
S_{BI} = -a \int d^{d+1} r \sqrt{m^2 c^2 - (\partial_\mu \theta )^2}. \label{eqn:BI}
\end{equation}
In non-relativistic limit, this model agrees with the Chaplygin gas.
These two models are equivalent and are transformed to the other by exact transformation 
\cite{Jackiw-Polychronakos}.
Several authors have given classical solutions of these two models \cite{Hoppe} $\sim$ \cite{Gibbons}.
In this paper we especially pay attention to the static solutions.
This is worthwhile because, if we find one static solution, 
we get the various time dependent solutions by combining Galileo boost and FDT.
There is another reason of considering the static solution.
The essence of these Chaplygin and Born-Infeld model is 
the theory of minimal surface or saying Plateau problem, 
and its typical form is the static equation. 
From this correspondence, the time dependent equations can also be rewritten 
into static form by using some ansatz.
The static equation has special symmetry which we call generalized scale transformation.
On this symmetry we discuss its geometrical meaning in section 2.
In section 3 we discuss the solution of static equation, 
and we show that there are some new non trivial solutions as minimal surfaces. 
In section 4 we discuss the time dependent solution in the sense of Plateau problem.
In section 5 the relation to Hamiltonian-BRST formalism is discussed.
In section 6 conclusion is given.

\section{Geometry and Symmetry 
\newline for Chaplygin gas}

     It is easy to see that d-dimensional Chaplygin gas has the d+2 dimensional Poincar\'e 
symmetry if we start from d-brane theory. 
There is another generalized scaling symmetry, which we will see in the followings.
Let us start from the Chaplygin gas equation with only $\theta $ variable.
\begin{equation}
\frac{\partial }{\partial t} \frac{1}{\sqrt{\dot{\theta }+ \frac{(\nabla \theta )^2}{2m}}}  + 
\nabla \cdot ( \frac{\nabla \theta /m}{\sqrt{\dot{\theta }+ 
\frac{(\nabla \theta )^2}{2m}}}) = 0. \label{eqn:motion}
\end{equation}
If we consider the time independent solution, the equation takes quite simple form, 
such as,
\begin{equation}
\nabla \cdot ( \frac{\nabla \theta}{\sqrt{(\nabla \theta )^2}}) = 0. \label{eqn:static1}
\end{equation}
The same equation can be obtained by taking the massless limit.
This equation means that the surface $\theta (r^1, r^2, \cdots , r^d) = const.$ 
has zero extrinsic mean curvature, where $r^1, r^2, \cdots , r^d$ are Cartesian 
coordinates on brane.  We will show this connection in the following.
Let us introduce the d-1 dimensional hypersurface defined by 
$\theta (r^1, r^2, \cdots , r^d) = 0$ in d-dimensional Euclidean space (d-brane).
We can introduce d-1 dimensional coordinates $(q^1, q^2, \cdots , q^{d-1})$ 
on that surface with its induced metric as
\begin{equation}
g_{ij} \equiv  \frac{\partial \vec{r}}{\partial q^i} \cdot \frac{\partial \vec{r}}{\partial q^j},
\end{equation}
where $\{i,j,\cdots \} $ run from 1 to d-1.
Inverse metric is defined as $g^{ij} $.
Further we introduce another coordinate $q^d$ normal to that surface.
The surface is then specified by $q^d =0$.
The set $\{q^1, \cdots, q^d \}$ construct the curvilinear coordinate in d-dimensional space.
We use early Latin indices $\{a, b, \cdots \}$
for the set of coordinates from $q^1$ to $q^d$ like $q^a,q^b, \cdots$, and 
large Latin indices $\{A, B, \cdots \}$
for the set of coordinates from $r^1$ to $r^d$ like $r^A,r^B, \cdots$.
Then the metric in this curvilinear d-dimensional space on the surface ($ q^d =0$) is given as
 \[ g_{ab} = \left(  
       \begin{array}{c|c}
       g_{ij} & 0\\
       \hline
       0 & g_{dd}
       \end{array}
 \right), \]
 \[ g^{ab} = \left(  
       \begin{array}{c|c}
       g^{ij} & 0\\
       \hline
       0 & g^{dd}
       \end{array}
 \right). \]
Now we define extrinsic curvature by using metric on surface, and 
normal unit vector $\vec{n}$ to the surface.
The extrinsic curvature is defined as
\begin{equation}
\kappa _{ij} \equiv -\frac{\partial \vec{n}}{\partial q^i} \cdot 
\frac{\partial \vec{r}}{\partial q^j},
\end{equation}
and extrinsic mean curvature is given as
\begin{equation}
\kappa \equiv g^{ij} \kappa_{ij}.
\end{equation}
Note that $\vec{n}$ is defined only on surface and it does not depend on $q^d$.
If we transport the normal unit vector on surface, its direction changes, 
and its difference projected to the surface is the extrinsic curvature 
\cite{ogawa-fujii}.
Then we can show that extrinsic mean curvature is equal to $-div(\vec{n})$.
\begin{eqnarray}
\kappa & \equiv & g^{ij} \kappa_{ij} = - g^{ij} \frac{\partial \vec{n}}{\partial q^i} \cdot 
\frac{\partial \vec{r}}{\partial q^j}, \nonumber \\
& = & - g^{ab} \frac{\partial \vec{n}}{\partial q^a} \cdot 
\frac{\partial \vec{r}}{\partial q^b} =
 -\frac{\partial q^a}{\partial r^A} \frac{\partial q^b}{\partial r^A} 
\frac{\partial n^B}{\partial q^a} \frac{\partial r^B}{\partial q^b} \nonumber \\
& = & -\frac{\partial n^A}{\partial r^A}  \equiv -div(\vec{n}).
\end{eqnarray}
Therefore our static equation for Chaplygin gas means
\begin{equation}
\kappa =0,  \label{eqn:static2}
\end{equation}
which is similar to Einstein equation in vacuum $R=0$.
There is deep relation between intrinsic and extrinsic curvature \cite{ogawa-fujii}
\begin{equation}
R = \kappa ^2 - \kappa _{ij} \kappa ^{ij}.
\end{equation}
The surfaces satisfying $\kappa =0$ are called ``minimal surface", 
and various solutions are known.
These solutions are discussed in next section.
Here we clarify the symmetry of equation (\ref{eqn:static1}).
It is trivial that this equation is invariant under
\begin{equation}
\theta  \to \theta' = F(\theta ), \label{eqn:symmetry}
\end{equation}
where $F(x)$ is any real function satisfying $dF/dx \neq 0$. 
So if $\theta $ is the solution, $F(\theta )$ is also the solution.
This point is also noted in \cite{Gibbons}.
We call this symmetry as generalized scale invariance.
Under this transformation, action is not invariant but the equation of motion is invariant.
This kind of fact occurs usually in scale transformation \cite{scale}, 
and scale transformation corresponds to a special choice of $F$. 
Note that we can not use Noether theorem 
because it is not the symmetry of action.
We can show another equation of motion, 
which has time-dependence and invariance under this transformation.
\begin{equation}
V = \kappa +const. ~~~  V=\frac{\dot{\theta }}{\mid \nabla \theta \mid}, \label{eqn:mean}
\end{equation}
where $V$ is the velocity of growing surface $\theta (\vec{r},t)=0$.
This equation is called as mean curvature flow equation,
 and plays important role in the theory of crystal growth. 
This is found by Ohta, Jasnow, and Kawasaki in '82 \cite{Ohta-Jasnow-Kawasaki}, and
 discussed by several authors \cite{Gurtin},\cite{Yokoyama-Sekerka},\cite{Wettlaufer}.
 In this theory, physical quantity is not $\theta$ itself, 
 but is the surface defined by $\theta =0$. 
 Therefore the symmetry (\ref{eqn:symmetry}) is natural if we fix $F(x)$ to satisfy $F(0)=0$.
 (This can be done without loss of generality, 
 if we pull out another trivial symmetry $\theta ' = \theta  + Const.$.)
 Only in static case this theory is the same as ours. 
This symmetry is important to solve time dependent equation (\ref{eqn:motion}).
 Starting from one static solution, we have infinitely many static solutions by transformation $F$.
 Then we boost it to obtain the time dependent solutions, 
 and change it in non trivial way by using FDT. 
 The time-dependent solutions obtained this way depend on the choice of function $F$.
 In this sense,  this symmetry remains in the time-dependent solution 
 due to the Galileo invariance of the theory.
 So it may be possible to say, 
 this generalized scale symmetry is hidden in the theory of Chaplygin gas.
 Then we can give a conjecture that the mean curvature flow equation (\ref{eqn:mean}) 
 has possibility to transform into (\ref{eqn:motion}).

      In addition, if we take the massless limit for the Born-Infeld model, 
 this symmetry clearly appears as we see from its time-dependent equation of motion.
 \begin{equation}
 \mbox{Born-Infeld model}  \stackrel{m \to 0}  {\longrightarrow } \partial^\mu 
 (\frac{\partial _\mu  \theta }{\sqrt{(\partial_{\alpha}  \theta )^2}}) =0.
  \end{equation}

\section{Solution of $\kappa =0$}

\subsection{standard consideration}
Let us construct the classical solution for static Chaplygin gas (\ref{eqn:static1})
 by using separation of variables method.
 This equation can be written in the form.
 \begin{equation}
 G^{ijkl}\partial_i\theta \cdot \partial_j\theta \cdot \partial_k \partial_l \theta =0,~~~ 
  G^{ijkl}\equiv \delta^{ij}\delta^{kl}-\delta^{ik}\delta^{jl}.
 \end{equation}
 We look for special solution of the form:
 \begin{equation}
 \theta(r^1, r^2, \cdots r^d) = \theta^{(1)}(r^1)\cdot \theta^{(2)}(r^2)\cdots \theta^{(d)}(r^d).
 \end{equation}
 Putting this form into the equation, we obtain
 \begin{equation}
 \sum_{i \neq j} [(\frac{\partial_i \theta^{(i)}}{\theta^{(i)}})^2 \frac{\partial^2_j 
 \theta^{(j)}}{\theta^{(j)}} 
  -  (\frac{\partial_i \theta^{(i)}}{\theta^{(i)}})^2 
 (\frac{\partial_j \theta^{(j)}}{\theta^{(j)}})^2 ] =0.
 \end{equation}
 Introducing the new variable by
 \begin{equation}
 Z^i \equiv \log \theta^{(i)},
 \end{equation}
 we write our equation as
 \begin{equation}
 \sum_{i>j} f_{ij}(r^i, r^j) = 0, ~~~ f_{ij} \equiv[(\partial_i Z^i)^2 \partial^2_j Z^j + 
 (\partial_j Z^j)^2 \partial^2_i Z^i ].
 \end{equation}
 Here we assume further that all the $f_{ij}$'s vanish.\\
 Then we get
 \begin{equation}
  \partial^2_j Z^j = \lambda_j (\partial_j Z^j)^2 , ~~~ \lambda_j + \lambda_i =0.
 \end{equation}
 $Z^i$ satisfies the same equation.
 The solution is given by
 \begin{eqnarray}
  ~~~ Z^j &=& A_j r^j + B_j,\\
  \mbox{For} ~ \lambda = 0, ~~~ \mbox{and}~~~~ && \nonumber\\
  ~~~ Z^j &=& C_j -\frac{1}{\lambda_j}\log \mid r^j - R^j \mid,\\
  \mbox{For} ~ \lambda \neq 0,~~~~~~~~~~~~ && \nonumber
 \end{eqnarray}
 where $A,B,C,R$ are the integration constants.\\
 Therefore we have 
 \begin{equation}
 \theta^{(j)}= \exp[A_j r^j + B_j], 
 ~~~ \theta (r^1,r^2,\cdots r^d) = F(\vec{A}\cdot \vec{r}+ B),
 \end{equation}
 in the case of $\lambda=0$.
 where $F$ is any function, and $\vec{A}, B$ are constants.
 This solution has been obtained by Jackiw and Polychronakos \cite{Jackiw-Polychronakos}.
 Here we have another new solution.
In the case of $\lambda \neq 0$, $\lambda_i + \lambda_j =0$ 
should be hold for any pair (i,j) which is possible only when d=2 (membrane).
Then we have the solution
\begin{equation}
\theta(x,y) = C \cdot (\frac{y-y_0}{x-x_0})^\alpha,
\end{equation}
where $C, x_0, y_0, \alpha$ are constants.\\
Using the generalized scale transformation, this can be extended into
\begin{equation}
\theta(x,y) = F(\frac{y-y_0}{x-x_0}).
\end{equation}
This is the any function of rotation angle centered at $(x_0, y_0)$.
Therefore the normal vector for the surface $\theta=const.$ constructs 
vortex like vector field. 
As this solution has singularity at the origin, we should consider the membrane with finite area 
and appropriate boundary condition, where the origin is taken to be outside.

This solution is related to the higher dimensional solution: helicoid by choosing the
 form of F.
Let us define 
\[
\alpha \equiv a z + b + F( \frac{y-y_0}{x-x_0} ).
\]
For $\alpha$ to satisfy the static equation (\ref{eqn:static1}) again, we need the relation
\[
\nabla \cdot \frac{F'(\phi) \nabla \phi}{\sqrt{a^2 + F'(\phi)^2 (\nabla \phi)^2}} = 0,~~~\phi \equiv \frac{y-y_0}{x-x_0}.
\]
Then l.h.s has the following form.
\begin{eqnarray}
&&-\partial_x \frac{y-y_0}{\sqrt{(a^2 (x-x_0)^4/F'^2) + r^2}} + 
     \partial_y  \frac{x-x_0}{\sqrt{(a^2 (x-x_0)^4/F'^2) + r^2}},\nonumber \\
&& r^2 \equiv (x-x_0)^2 + (y-y_0)^2. \nonumber
\end{eqnarray}
If the denominator is function of only $r$, sum of these terms vanishes.
This occurs when $F'(\phi) = (x-x_0)^2 f(r)$ is satisfied. 
From dimensional analysis, $f= const/r^2$ holds. 
This gives
\[
  F'(\phi) = const. \times \frac{1}{1+\phi^2},
\]
which leads to $F(\phi)=\arctan (\phi)$ up to multiplicative constant.
So we come to 3 dimensional solution known as helicoid: $\alpha =0$,
\begin{equation}
\alpha \equiv az+b + \arctan(\frac{y-y_0}{x-x_0}).
\end{equation}

First two static solutions are not surprising in the view point of geometry,
because the surface $\theta =const.$ should have zero mean curvature.
The first solution gives the flat surface which is normal to $\vec{A}$, and the surface given 
by second solution is the half straight line, which is flat except the terminal.
In d=2 the surface $\theta =const.$ gives the 1 dimensional line, which should be straight 
from geometrical requirement. 
In this sense these two solutions are general ones up to generalized scale invariance for d=2.
For higher dimensions we have other solutions as we have seen one d=3 solution.
In fact some minimal surfaces are known at d=3, and they are explained in next subsection.

\subsection{d=3 Static solutions as minimal surfaces}
If we find the minimal surface as $G(r^1,r^2, \cdots r^d) =0$, we have
\begin{equation}
\kappa[G] \equiv \nabla \cdot ( \frac{\nabla G}{\sqrt{(\nabla G)^2}}) = 0 ~~~~ 
on ~G=0.\label{eqn:minimal}
\end{equation}
The key point is that $l.h.s.$ vanishes only on the surface $G=0$.
This is the difference to our solutions where the mean curvature vanishes everywhere 
even for $G \neq 0$.
In a special case where we write the function $G$ as $G \equiv r^d-f(r^1, r^2, \cdots r^{d-1})$,
$\kappa[G]$ does not depend on $r^d$ at all.
Then $\kappa[G] =0$ is satisfied without condition $G = 0$. 
Therefore this minimal surface can be the solution of our problem.
Our static solution for Chaplygin gas is given as 
\begin{equation}
\theta(\vec{r}) = F( G(\vec{r}-\vec{R}) ),
\end{equation}
where F is any analytic function, and $\vec{R}$ is the constant.
In the following we show some minimal surfaces.
(We have non trivial minimal surfaces at $d \geq 3$. Because $\kappa =0$ does not mean flat.
For any point on hypersurface G=0, we can take d-1 orthogonal tangent vectors.
The mean curvature is the sum of extrinsic curvatures for each directions.)

Let us here discuss d=3 case which is well studied as Plateau problem.
In this case some non trivial minimal surfaces are known.\cite{Kobayashi}\\
One example is the catenoid given as
\begin{equation}
 \sqrt{x^2 + y^2} - a \cosh\frac{z}{a} = 0,
\end{equation}
where $a$ is the positive constant.
Another example is the right helicoid which we have seen before.
\begin{equation}
x= u \cos v, ~~ y= u \sin v, ~~ z= av +b; ~~~ y - x \tan(\frac{z-b}{a}) =0,
\end{equation}
where $u$ and $v$ are the real parameters, $a$ and $b$ are the constants.
We present two new examples.\\
First one is Scherk's minimal surface, which is given as
\begin{equation}
e^z \cos x - \cos y =0.
\end{equation}
Second one is Enneper's minimal surface, which is given as
\begin{equation}
x= 3u + 3uv^2 -u^3,~~~ y= v^3 -3v -3u^2v, ~~~ z= 3(u^2 - v^2).
\end{equation}
\\
For right helicoid we have solutions as Chaplygin gas,
\begin{eqnarray}
\theta(x,y,z) &=& F( y-y_0 -(x-x_0)\tan(az+b)),\nonumber \\
 or, ~~~\theta(x,y,z) &=& F( x-x_0 -(y-y_0)/ \tan(az+b)),\nonumber \\
 or, ~~~\theta(x,y,z) &=& F(z-z_0 - \arctan(\frac{y-y_0}{x-x_0})).
\end{eqnarray}
For Scherk's minimal surface,
\begin{eqnarray}
\theta(x,y,z)&=& F( z-z_0 -\log \mid \frac{\cos(y-y_0)}{\cos(x-x_0)}\mid),\nonumber \\
 or, ~~~\theta(x,y,z)&=& F( x-x_0 -\arccos(e^{\pm (z-z_0)}\cos(y-y_0))).
\end{eqnarray}
For catenoid,
\begin{eqnarray}
\theta(x,y,z)&=& F( x-x_0 \pm \sqrt{a^2 \cosh^2(\frac{z-z_0}{a})-(y-y_0)^2 }),\nonumber \\
 or, ~~~\theta(x,y,z)&=& F( z-z_0 - a \cosh^{-1}(\frac{\sqrt{(x-x_0)^2 + (y-y_0)^2}}{a})).
\end{eqnarray}

  ~~~But for Enneper's one we can not simply write it in the form like \\
$G(x,y,z) =0$. 
So it is still not clear at this stage. 
We just only put the extrinsic curvature and induced metric on this surface.
\[ g_{ij} = 9( 1 + u^2 + v^2 )^2 \left(  
       \begin{array}{c|c}
       1& 0\\
       \hline
       0 & 1
       \end{array}
 \right), \]
 \[ \kappa_{ij} = 6 \left(  
       \begin{array}{c|c}
       1 & 0\\
       \hline
       0 & -1
       \end{array}
 \right). \]
which leads to $\kappa =0$ on surface.

From above consideration, 
at least we have three kinds of new non trivial static solutions for Chaplygin gas at d=3.
For catenoid and helicoid solutions, 
they are also discussed by Gibbons \cite{Gibbons} for static Born-Infeld model.
For these static solutions, time dependence can be introduced by Galileo symmetry and then changed by FDT.
In this way we can construct d=3 solutions as many as the minimal surfaces.
It should be stressed that catenoid solution can be easily extended to any dimensions by using ansatz
\[ \theta = r^{d} - f(\bar{r}), ~~~ \bar{r}^2 = (r^1)^2 + (r^2)^2 + \cdots + (r^{d-1})^2. \]

Some of minimal surfaces are determined by minimization of area with fixed boundary.
This is well known as Plateau problem.
If we give the boundary as closed loop, 
the shape of soap membrane can be determined by minimization of area.
This is just the minimal surface: $\kappa =0$ everywhere on surface.
(The opposite is not necessarily true.)
On this point we give a bit explanation.
Let us consider the surface as $z=f(x,y)$ with some fixed boundary C (closed loop).
Then the surface with minimal area with fixed boundary is given by
\begin{equation}
 \delta \Gamma[f]=0,  ~~~\Gamma[f] \equiv \int \int_D dxdy \sqrt{1+ (\nabla f)^2},\label{eqn:plateau}
\end{equation}
where D is the region closed by C projected to x-y plane, and $\nabla$ is for x and y.
The reason is simple. The infinitesimal area of surface is 
$$d\Gamma = \frac{dx dy}{n_z},$$
where $\vec{n}$ is the normal unit vector to that surface. This is given by
$$ \vec{n}=\frac{\nabla \theta}{\mid \nabla \theta \mid}=(\frac{-\partial_x f}
{\sqrt{1+ (\nabla f)^2}}, \frac{-\partial_y f}{\sqrt{1+ (\nabla f)^2}}, \frac{1}{\sqrt{1+ (\nabla f)^2}}).$$
Here we write $\theta \equiv z-f(x,y)$.
The variation of $\Gamma$ gives equation
\begin{equation}
\frac{\partial}{\partial x}(\frac{\partial_x f}{\sqrt{1+(\nabla f)^2}}) + 
\frac{\partial}{\partial y}(\frac{\partial_y f}{\sqrt{1+(\nabla f)^2}}) =0.\label{eqn:plateau2}
\end{equation}
This is rewritten as
\begin{equation}
\nabla \cdot (\frac{\nabla \theta}{\mid \nabla \theta \mid}) =0.
\end{equation}
This means the surface is minimal: extrinsic mean curvature is vanishing.
So we should finally say that, finding the solution of d-dimensional static Chaplygin gas 
 is similar to finding the d-1 dimensional minimal area surface in d-dimensional space.
The relation to Plateau problem with our time dependent d-brane theory 
will be discussed in the next section.

\section{Time dependent solutions
\newline
 and Plateau problem}

We show the time dependent solutions and their connections with Plateau problems in this section.
The Plateau action, $\Gamma$ in equation (\ref{eqn:plateau}) has a good similarity with time dependent 
action for Chaplygin gas and Born-Infeld model.\\
First we consider the Chaplygin gas with action (\ref{eqn:chaplygin}).
We look for the solution in the form:
\begin{equation}
 \theta = t - \sqrt{2m}~f(\vec{r}).
\end{equation}
Then the equation for $f(\vec{r})$ is the same as the equation (\ref{eqn:plateau2}). 
This is the minimum area problem for hypersurface $z=f(\vec{r})$, 
 or saying Plateau problem in d+1 dimensional space.
The boundary condition is just the one for d-brane.
Note that in this case there is no generalized scale symmetry for $f$.
The dimension of surface is now $d$ but not $d-1$.\\
Therefore if we have d-dimensional minimal surface, 
we can construct the solution of d-spatial dimension as above.
We should remark that this time dependent solution is time-independent in the sense of fluid mechanics.
Because the velocity $\vec{v} = \nabla \theta$, and the density $\rho$ which is obtained from $\dot{\theta}$ and 
$\nabla \theta$, are really time independent. To obtain the explicit time dependence, FDT or Galileo boost is necessary.

Next we come to the Born-Infeld model with rescaled action of (\ref{eqn:BI}).
\begin{equation}
S_{BI} = - \int dtd^dr \sqrt{1-(\partial_\mu \theta)^2}. \label{eqn:BI2}
\end{equation}
For this action, if we take ansatz like $\theta = t- \beta(\vec{r})$, $\beta$ satisfies (\ref{eqn:static1})
 and $\beta=0$ gives the d-1 dimensional minimal surface in d-dimensional space. 
So we come again to Plateau problem. 
This ansatz was discussed by Hoppe \cite{Hoppe2} in d=2 case.

In this point, let us show another observation of second different problems.
If we compare (\ref{eqn:BI2}) with action (\ref{eqn:plateau}), 
this is just the relativistic version of Plateau problem, 
and we should consider the d+1 dimensional minimal surface 
as $z=\theta(t,r^1,r^2, \cdots r^d)$ in d+2 dimensional Minkowskian space, 
in which the essence is the same as original Nambu-Goto action: minimization of world surface.
It is possible to rewrite this Born-Infeld action into the d+2 dimensional 
form as we have discussed already. Let $x^A = \{t, r^1, r^2, \cdots ,r^d, z \}$ and 
taking metric $\eta_{AB} = (1,-1,-1, \cdots, -1)$. 
Then for new variable $\alpha$,
\begin{equation}
\alpha \equiv z-\theta(t,r^1,r^2,\cdots ,r^d),
\end{equation}
the action
\begin{equation}
S_{BI} = - \int dt d^dr dz \sqrt{-\eta^{AB} \partial_A \alpha \cdot \partial_B \alpha},
\end{equation}
gives the same theory as Born-Infeld model.
The equation of motion is
\begin{equation}
\frac{\partial}{\partial x^A} 
(\frac{\partial^A \alpha}{\sqrt{(\partial_B \alpha)^2}}) =0. \label{eqn:minkowski}
\end{equation}

This means that $\alpha =0$ is the d+1 dimensional ``Minkowskian minimal surface" just like 
(\ref{eqn:static1}).
The generalized scale symmetry appears here, 
but the``gauge choice" $\alpha = z-\theta(t,r^1,r^2,\cdots ,r^d)$ 
gives the Born-Infeld model.

   The property of d+1 dimensional Minkowskian minimal surface given by equation (\ref{eqn:minkowski}) 
with weaker condition $\alpha=0$ is discussed by several authors \cite{Hoppe3},\cite{Graustein},\cite{Gibbons}
, though they did not discuss the connection with Plateau problem.

There appears the solution constructed by elliptic Weierstrass function forming the BIonic crystal for d=2. 
Here we give new different special solutions, based on quite simple observation.
By comparing (\ref{eqn:minkowski}) with (\ref{eqn:static1}), 
the difference appears only on signature for the square of time derivative.
So if we take analytic continuation for minimal surfaces, we obtain the solution for $\alpha$.
For example some solutions of Plateau problem in d=3 is known, and 
if we change one variable to imaginary time: for example $y=it$, 
the ``Minkowskian surface": $z = f(x,t)$ is obtained. 
Then $\alpha = z-f(x,t)$ satisfies equation (\ref{eqn:minkowski}).
Therefore $f$ is the solution of Born-Infeld model. 

Let us see some d=1 solutions of (\ref{eqn:minkowski}) constructed from 
3 typical minimal surfaces by using the analytic continuation.
First from catenoid,
\begin{equation}
\theta_{BI} = \sqrt{\omega^{-2} \cos^2(\omega(t-t_0))-(x-x_0)^2 }.
\end{equation}

From right helicoid,
\begin{equation}
\theta_{BI} = (x-x_0) \tan(\omega t+b).
\end{equation}

From Scherk surface,
\begin{equation}
\theta_{BI}= \log \mid \frac{\cosh(t-t_0)}{\cos (x-x_0)} \mid.
\end{equation}

By using the transformation from Born-Infeld model to Chaplygin gas, 
the solutions for d=1 Chaplygin gas are obtained.
They should be some of general solutions given by Bazeia \cite{Bazeia}.
In this way there will be path between string in (2,1) dimension and Plateau problem in d=3.
Inversely starting from general solution for d=1 Chaplygin gas, 
we might obtain the general solution for Plateau problem in d=3.
Therefore we can obtain (d-1,1) Born-Infeld solutions 
from d-dimensional minimal surface in d+1 dimensional space.

\section{A comment on the relation to \newline  Hamiltonian-BRST formalism}
The treatment of constraint problem for the membrane theory has been discussed for a long time. 
The problem is due to its property of open-algebra, that is, 
the field dependence of structure constant for the Poisson brackets between first class constraints. 
The Hamiltonian BFV-BRST formalism of path-integral has reached to the result on '83 
given by M.Henneaux \cite{FV} \cite{Henneaux}. 
For the  path-integral in configuration space, Fujikawa and Kubo have given 
another way \cite{Fujikawa}. 
These two methods are equivalent and constructed on the basis of explicit 
covariant gauge for target space. For the equivalence see appendix.
In both cases, there appears 4-ghost term, 
and it seems usual for general d-brane to have such ghost terms.  
But this is not true for other non-covariant gauge.
The starting idea on Hamiltonian BFV-BRST formalism is to introduce the covariant gauge 
for the first ordered (Hamiltonian formed) path-integral formula.
 (I mean the covariant gauge as the one including the time derivative of auxiliary field.) 
In this sense even if we take the gauge function equals to zero, 
gauge is fixed already to take into account the time derivative of auxiliary field 
as gauge condition in this formalism \cite{FV},\cite{Ogawa}.
So if we work with other gauge, this framework fails.

 In the framework given by Hoppe and Bordemann, the gauge condition was not 
 so clear in the way of Dirac method for the diffeomorphism symmetry.
 But the quite simple gauge condition $X^i = r^i$ to fix 
 that local symmetry gives the same result as the ones early given by Hoppe and Bordemann.
 This is shown by Hoppe himself in \cite{Hoppe2}.
Use of this gauge condition with Light-cone gauge (Cartesian gauge) changes the action 
into the form of Chaplygin gas (Born-Infeld model), 
which takes canonical form completely as (\ref{eqn:chap}), and we have no ghosts in this gauge choice.
In fact if we work with Faddeev-Senjanovic formula, we can show it explicitly.
(even if we start from BFV- BRST framework, this is so. on this point see appendix.) 
We start from Nambu-Goto action for (d+1,1) Minkowskian target space ($X^0 \cdots X^{d+1}$) 
in (d,1) parameter space-time ($r^0 \equiv t, r^1, \cdots r^d $),
\begin{equation}
S= -\int d^{d+1}r ~ \sqrt{G}, ~~~ G \equiv (-1)^{d} det [\eta_{AB} ~ 
\frac{\partial X^A}{\partial r^\mu} \frac{\partial X^B}{\partial r^\nu} ],
\end{equation}
\begin{equation}
Z= \int {\cal D}X {\cal D}P \prod_{\mu} \delta(T_{\mu}) \prod_{\nu} \delta(\chi^{\nu}) 
\mid det \{T_\mu , \chi^\nu \}\mid \exp i \int d\tau d^dr ~ P_A \dot{X}^A ,
\end{equation}
where $T_\mu$ is the Hamiltonian-momentum constraint, and $\chi^\nu$ is the gauge condition.
Let us choose the Light-cone gauge
\begin{equation}
\chi^0 = \frac{1}{\sqrt{2}}(X^0 + X^{d+1})-t, ~~~ \chi^k = X^k - r^k ~~(k=1 \sim d),
\end{equation}
with the definition $ \theta \equiv \frac{1}{\sqrt{2}}(X^0 - X^{d+1})$. Then we obtain
\begin{equation}
Z = \int {\cal D}\theta {\cal D}\rho \exp i \int dt d^dr[\theta ~ \dot{\rho} -
\frac{\rho}{2}(\nabla \theta)^2 - \frac{1}{2\rho}],
\end{equation}
where $\rho \equiv \frac{1}{\sqrt{2}}(P_0 - P_{d+1})$.
If we take the Cartesian gauge
\begin{equation}
\chi^0 = X^0 - t, ~~~ \chi^k = X^k - r^k~~(k=1 \sim d),
\end{equation}
with the definition $ \theta \equiv X^{d+1}$, we obtain
\begin{equation}
Z = \int {\cal D}\theta {\cal D}\rho \exp i \int d\tau d^dr[\theta ~ \dot{\rho} -
 \sqrt{\rho^2 - 1} \sqrt{1+ (\nabla \theta)^2}],
\end{equation}
where $\rho \equiv P_{d+1}$.
These are the Chaplygin gas and Born-Infeld model.

\section{Conclusion}
We have discussed on the classical solutions for Chaplygin gas as d-branes.
The static equation for this theory has the geometrical meaning 
as vanishing extrinsic mean curvature.
This means that the $\theta$ field extended on d-brane is considered as the set of contour lines, 
and these lines, or saying these surfaces, have the vanishing mean curvature.
Such surfaces of vanishing mean curature are known as minimal surfaces.
As a result if we find one d-dimensional minimal surface in d+1 dimensional space,
 we have a static d+1 dimensional solution for Chaplygin gas. 
In this way we have shown some examples of static solutions for Chaplygin gas from minimal surfaces.
Furthermore if we fix the time dependence as $\theta = t - \sqrt{2m} f(\vec{r})$,
 we obtain the solutions of d-spatial dimension from d-dimensional minimal surface in d+1 dimensional space.
For the Born-Infeld model, this is just the relativistic extension of Plateau problem and 
this interpretation says the same meaning as original Nambu-Goto action:
 minimization of area for world hypersurface.
Solving this Born-Infeld model as Minkowskian Plateau problem will be interesting.
This is not only for its intuitive form, but also for its connection with usual Plateau problem 
by analytic continuation.
For example the general solutions for Plateau problem in d=3 might be obtained from 
the general solution given by Bazeia \cite{Bazeia} for string in (2,1) space time.
In many cases our two models are closely  related to the Plateau problem.
If we find d-dimensional minimal surface in d+1 dimensional space,
we obtain the solutions of Chaplygin gas for d and d+1 spatial dimension,
and the solutions of Born-Infeld model for d+1 and d-1 spatial dimension.
Since these two models are related by exact transformation, 
we obtain the solutions of d-1,d,d+1 spatial dimensions from d-dimensional minimal surfaces.
The theorem for the existence of solution for Plateau problem is related to
 the mapping theorem of Riemann in the theory of conformal mapping \cite{Courant}.
To study this point as physics may be interesting open problem.

\section{Appendix}
For the membrane theory, Hamiltonian BRST formalism takes the form
\begin{equation}
S_{eff}=\int dt [P_A \dot{X^A} + \dot{C}^\mu \bar{C}_\mu + \dot{\eta}^\mu \bar{\eta}_\mu+ 
\Pi_\mu \dot{N}^\mu + \{Q_B, ~ \bar{C}_\mu N^\mu + \bar{\eta}_\mu \chi^\mu \}],
\end{equation}
\begin{equation}
Q_B = \eta^\mu \Pi_\mu + \Omega,~~~ \Omega \equiv C^\mu T_\mu + C^\mu C^\nu \:
^{(1)}U^{\lambda}_{\nu \mu} \bar{C}_\lambda + C^\mu C^\nu C^\rho 
\:^{(2)}U^{\lambda \sigma}_{\rho \nu \mu} \bar{C}_\sigma \bar{C}_\lambda,
\end{equation}
where $C,\bar{C},\eta,\bar{\eta}$ are Grassmannian odd ghost fields, $\Pi, N$ are the auxiliary fields, 
$\chi$ is the gauge fixing function, and $T_\mu$ is the Hamiltonian and momentum constraint. 
$\:^{(1)}U$ is the structure constant for the Poisson brackets between 1st class constraints,
 and $\:^{(2)}U$ is introduced to obtain the nilpotency of $Q_B$. It is proved that 
 the theory does not depend on the choice of  $\chi$ \cite{FV},\cite{Henneaux}.
Even if we take $\chi = 0$, gauge is fixed already since gauge condition is $\dot{N}+ \chi =0$ \cite{Ogawa}.
2-dimensional spatial integration is included in the contraction of indices. The indices 
$\mu,\nu,\rho, \cdots $ run from 0 to 2 with spatial integration. All the variables are
 forming the canonical sets. For the path-integral
\begin{equation}
Z = \int {\cal D}X {\cal D}P {\cal D}\bar{C} {\cal D}C {\cal D}\bar{\eta}
 {\cal D}\eta {\cal D}N {\cal D}\Pi  ~ \exp [i S_{eff}],
\end{equation}
we change the integral variables like $\bar{\eta} \to \epsilon \bar{\eta} 
,~ \Pi \to \epsilon \Pi, ~ \chi \to \chi/\epsilon$ then the path-integral measure does not change.
Then we take limit $\epsilon \to 0$ which reduces the theory into
\begin{equation}
S_{eff}=\int dt [P_A \dot{X^A} + \dot{C}^\mu \bar{C}_\mu - \eta^\mu \bar{C}_\mu - 
\Pi_\mu \chi^\mu + \{\Omega, ~ \bar{C}_\mu N^\mu + \bar{\eta}_\mu \chi^\mu \}].\label{eqn:modify}
\end{equation} 
If we choose $\chi=\chi(X,P)$, $\eta$ integration can be performed and to obtain 
$\bar{C}=0$. In that case, the theory reduces to
\begin{equation}
S_{eff}=\int dt [P_A \dot{X^A} - \Pi_\mu \chi^\mu - T_\mu N^\nu 
+ C^\mu \{T_\mu , \chi^\nu \}\bar{\eta}_\nu ].
\end{equation}
This is clearly equals to the Faddeev-Senjanovic formula.
In this way if we work with gauge like Light-cone or Cartesian, we come to the
 Faddeev-Senjanovic formulation even if we start from BFV BRST formalism.
But for general choice of gauge, this is no longer hold.
Starting from (\ref{eqn:modify}), we have

\begin{eqnarray}
S_{eff}&=&\int dt [P_A \dot{X^A} + \dot{C}^\mu \bar{C}_\mu - \eta^\mu \bar{C}_\mu - 
\Pi_\mu \chi^\mu + \{\Omega, ~ \bar{C}_\mu N^\mu + \bar{\eta}_\mu \chi^\mu \}] \nonumber \\
&=& \int dt [P_A \dot{X^A} + \dot{C}^\mu \bar{C}_\mu  + \delta_{B}(N^\mu \bar{C}_\mu 
+ \bar{\eta}_\mu \chi^\mu) ],\label{eqn:modify2}
\end{eqnarray}
where,
\begin{eqnarray}
\delta_{B}F(X,P,C,\bar{C}) &\equiv& \{ \Omega, F(X,P,C,\bar{C})\},\nonumber\\
\delta_{B}\Pi_\mu &=& 0, \nonumber\\
\delta_{B}\eta^\mu &=& 0, \nonumber\\
\delta_{B}N^\mu &=& \eta^\mu, \nonumber\\
\delta_{B}\bar{\eta}_\mu &=& -\Pi_\mu. \nonumber\\
\end{eqnarray}
These defined BRST transformations are proved to be nilpotent, and it is easily seen that
 $ \Pi,\eta, N, \bar{\eta} $ are forming the BRST quartet.
The theory starting from the action (\ref{eqn:modify2}) is given by M.Caicedo, A.Restuccia, 
and R.Torrealba and called modified BFV quantization \cite{Caicedo},
though they do not discuss on the reduction from Hamiltonian BRST formalism as above.
Then we take the gauge condition given by R.Torrealba and A.Restuccia,
\begin{equation}
\chi^0 = N^0 -1, ~~ \chi^1 = N^1, ~~ \chi^2 = N^2.
\end{equation}
$\eta, \bar{\eta}, \Pi,N $ integrations can be performed explicitly, and we obtain the form,
\begin{eqnarray}
Z &=& \int {\cal D}X {\cal D}P {\cal D}C {\cal D}\bar{C} ~~ \exp i\int dt [P_A \dot{X^A} + 
\dot{C}^\mu \bar{C}_\mu - H_{eff}], \nonumber \\
H_{eff} &=& T_0 + 2 C^\nu \:^{(1)}U^{\lambda}_{\nu 0} \bar{C}_\lambda +
 3 C^\nu C^\rho \:^{(2)}U^{\lambda \sigma}_{\rho \nu 0} \bar{C}_\sigma \bar{C}_\lambda.
\end{eqnarray}
Here we take the boundary condition that ghosts vanish at the boundary of membrane.
Because the Hamiltonian is quadratic for momentum,
the momentum integration can be performed explicitly and giving,
\begin{eqnarray}
S_{eff} &=& \int dt d^2r [ \frac{1}{2}\dot{X}^2 - \frac{1}{2}det[G_{kl}] +
 i\bar{C}_\mu \dot{C}^\mu - i\bar{C}_0 \partial_k C^k ], \nonumber \\
 && G_{kl} \equiv g_{kl} + i\bar{C}_k \partial_l C^0 + i \bar{C}_l \partial_k C^0,
\end{eqnarray}
where we have changed the variable $\bar{C} \to -i\bar{C}$.
This is Fujikawa and Kubo's formulation of membrane theory.
From the above consideration, we have shown various formulations 
are equivalent up to gauge choice and boundary condition for ghosts.
$$ \mbox{} $$
\noindent{\em Acknowledgement.}\\
The author would like to greatly thank Prof.R.Jackiw for giving him 
many information on this problem, valuable discussion and encouraging him. 
He thanks Prof.K.Ishikawa for the discussion and careful reading through the manuscript.
He also thanks Prof.Bazeia and Prof.Gibbons for important comments, and 
Prof.Yokoyama and Prof.Furukawa for informations on theory of crystal growth.

\end{document}